\documentclass[
    reprint,
    preprintnumbers,
    superscriptaddress,
    nofootinbib,
     amsmath,amssymb,
     aps,s
     prd,
    floatfix,
    ]{revtex4-2}

    \usepackage{graphicx}
    \usepackage[caption=false]{subfig}
    \usepackage[usenames,dvipsnames]{xcolor} 
    \usepackage[utf8]{inputenc}
    \usepackage{color}
    \usepackage{hyperref}
    \usepackage[normalem]{ulem} 
    \usepackage{physics}

    \usepackage{enumitem}
    \usepackage{comment}
    \usepackage{bm}

\graphicspath{{./}}

\hypersetup{
  colorlinks   = true, 
  urlcolor     = blue, 
  linkcolor    = blue, 
  citecolor   = blue 
}

\newcommand{\iu}{{i\mkern1mu}}
\newcommand{\msun}{{M_\odot}}

\begin{document}
\title{Revisiting the cosmic string origin of GW190521}
\author{Josu C. Aurrekoetxea}
\email{josu.aurrekoetxea@physics.ox.ac.uk}
\affiliation{Astrophysics, University of Oxford, Oxford OX1 3RH, United Kingdom}
\author{Charlie Hoy}
\email{charlie.hoy@port.ac.uk}
\affiliation{University of Portsmouth, Portsmouth, PO1 3FX, United Kingdom}
\author{Mark Hannam}
\affiliation{Gravity Exploration Institute, Cardiff University, Cardiff, United Kingdom}

\begin{abstract}

For the first time we analyse gravitational-wave strain data using waveforms constructed from strong gravity simulations of cosmic string loops collapsing to Schwarzschild black holes; a previously unconsidered source. Since the expected signal is dominated by a black-hole ringdown, it can mimic the observed gravitational waves from high-mass binary black hole mergers. To illustrate this, we consider GW190521, a short duration gravitational-wave event observed in the third LIGO--Virgo--KAGRA observing run. We show that describing this event as a collapsing cosmic string loop is favoured over previous cosmic string analyses by an approximate log Bayes factor of $22$. The binary black hole hypothesis is still preferred, mostly because the cosmic string remnant is non-spinning. It remains an open question whether a spinning remnant could form from loops with angular momentum, but if possible, it would likely bring into contention the binary black hole preference. Finally, we suggest that searches for ringdown-only waveforms would be a viable approach for identifying collapsing cosmic string events and estimating their event rate. This work opens up an important new direction for the cosmic-string and gravitational-wave communities

\end{abstract}
\maketitle

\textbf{\textit{Introduction.}}— The observation of gravitational-waves (GWs)~\cite{LIGOScientific:2021djp,Nitz:2021zwj,Venumadhav:2019lyq,Zackay:2019tzo,Zackay:2019btq} 
has paved the way to search for new physics. Cosmic strings \cite{Kibble:1976sj,Vilenkin:1981zs,Vilenkin:1981kz,Vilenkin:1984ib,Turok:1985tt,Hindmarsh:1994re} are a well-motivated example that naturally arise when the rapid cooling of the universe triggers a phase transition \cite{Vachaspati:1984dz,Jeannerot:2003qv,Copeland:2009ga}. Cosmic strings may manifest themselves through several channels, such as imprints via lensing on the Cosmic Microwave Background (CMB) \cite{Vilenkin:1984ea,Planck:2013mgr}, and a stochastic background of gravitational waves (SGWB) \cite{Vilenkin:1981bx,Vachaspati:1984gt,Allen:1991bk,Hogan:1984is,Allen:1991bk}, which is the total integrated power of incoherent GWs from all individual emissions that are too weak to be detected. The LIGO--Virgo--KAGRA (LVK) collaboration currently searches for the SGWB \cite{LIGOScientific:2021nrg}, and places constraints on the dimensionless string tension $G\mu/c^2$; a key property that sets their gravitational coupling strength and the energy scale of the phase transition, providing a unique link to the early universe. Localized coherent events of cosmic strings can also be searched for, if they are energetic enough to be directly detected.

To date, all observed GW signals are consistent with binary black hole (BBH) and/or neutron star mergers~\cite{LIGOScientific:2021djp,Nitz:2021zwj,Venumadhav:2019lyq,Zackay:2019tzo,Zackay:2019btq}, with signal-to-noise ratios that are mainly gained during the inspiral phase (around $20-50\mathrm{Hz}$). The absence or modification of this stage on the GW signal may 
serve as smoking gun for new physics. A candidate with these properties is GW190521 \cite{LIGOScientific:2020iuh,LIGOScientific:2020ufj}, which featured as a short transient in the LIGO--Virgo--KAGRA (LVK) detection pipelines due to its low frequency nature. The origin of this event has been extensively discussed in the literature as a massive black hole binary merger~\cite{Fishbach:2020qag,Nitz:2020mga,Estelles:2021jnz,Capano:2022zqm,Xu:2022zza,Sberna:2022qbn,Siegel:2023lxl}, eccentric encounter \cite{Romero-Shaw:2020thy,Gayathri:2020coq,Tagawa:2020jnc,Gamba:2021gap,Iglesias:2022xfc,Guo:2022ehk,Ramos-Buades:2023yhy,Morton:2023wxg} or new physics \cite{CalderonBustillo:2020fyi, Sakstein:2020axg,Clesse:2020ghq,Abedi:2021tti,CalderonBustillo:2022cja}. Another studied (and disfavoured) hypothesis of interest to this Letter is a cosmic string cusp \cite{LIGOScientific:2020ufj}, a gravitational-wave burst released when a fragment of a string doubles back on itself and moves at the speed of light. This waveform inherently lacks of a quasi-normal mode ringdown phase, as no black hole is formed during the process \cite{Damour:2000wa}.

\begin{figure*}[t!]
    \centering
    \includegraphics[width=\linewidth]{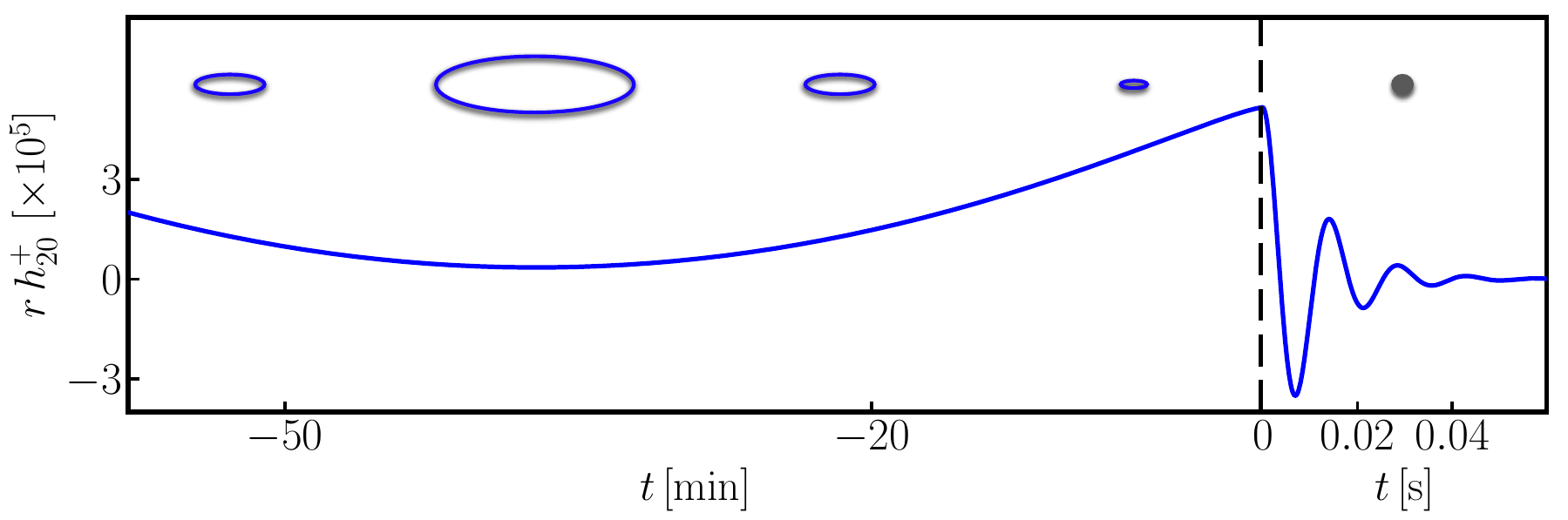}
    \caption{Gravitational waveform $h^+_{\ell m}$ of a collapsing cosmic string loop with radius $R_0=2.85\,\mathrm{AU}$ and  dimensionless string tension $G\mu/c^2=10^{-7}$. This corresponds to an initial mass of $M_0 \approx 181\msun$ that collapses to form a remnant black hole of mass $M\approx 175 \msun$ in the source frame, radiating $\approx 6\msun$ in gravitational waves. The vertical dashed line depicts the transition between the infall and black hole formation stages, where there is a sudden change in the frequency content of the signal.}
\label{fig:waveform}
\end{figure*}

In this Letter we revisit the cosmic string hypothesis of GW190521 using novel GW waveforms of cosmic string loops collapsing to black holes\footnote{We take the agnostic view of whether these circular loops are sufficiently abundant in nature and test the hypothesis by searching for their evidence in data.} \cite{Aurrekoetxea:2020tuw}. Guided by the first fully general relativistic field-theory simulations of Abelian Higgs strings, we construct an example waveform for a loop with dimensionless string tension $G\mu/c^2= 10^{-7}$ that collapses to form a black hole, radiating $3.4\%$ of its initial mass in gravitational waves, see Fig. \ref{fig:waveform}. At a fiducial distance $d_L = 4\,\mathrm{Gpc}$, this corresponds to a cosmic string loop with radius $R_0=2.85\,\mathrm{AU}$ observed with an inclination angle $\iota = 34^\circ$. We show that this collapsing cosmic string yields a signal-to-noise ratio $\rho \approx 12$, greatly improving on the cusp result ($\rho \approx 10$). This maps to an approximate log Bayes' factor of $\log \mathcal{B}^{\mathrm{loop}}_{\mathrm{cusp}} = 22$ in favour of the collapsing cosmic string over existing cusp results. Although the collapsing cosmic string is disfavoured relative to the BBH hypothesis ($\rho \approx 14$), a full Bayesian analysis that encompasses the full configuration space could potentially show comparable agreement. The purpose of our analysis is to show that while cusps are easily distinguishable from black-hole mergers, collapsing loops can be excellent mimickers. This makes it difficult to conclusively identify a high-mass binary black hole merger if collapsing cosmic string loops occur in nature.\\

\textbf{\textit{Collapsing string loop waveforms.}}— After cosmic strings form and the universe evolves, long strings self-intersect producing a network with abundant closed loops of a vast range of sizes \cite{Martins:1996jp,Ringeval:2005kr,Martins:2005es,Lorenz:2010sm,Blanco-Pillado:2011egf,Blanco-Pillado:2013qja,Allen:1990tv,Martins:2000cs}. These loops can sur oscillate and contribute to the SGWB \cite{Blanco-Pillado:2017oxo,NANOGrav:2023hvm,Ellis:2020ena,Blasi:2020mfx,Blanco-Pillado:2021ygr,Ringeval:2017eww} or collapse, process during which the loop may circularize\footnote{Details about their circularization timescales remain unknown as state-of-the-art network simulations do not incorporate the impact that gravitational backreaction has in the smoothing of the loops. See \cite{Chernoff:2018evo,Blanco-Pillado:2019nto} for promising work in the linearized limit.}. The linear mass density $\mu$ and radius $R_0$ of circular string loops set their rest mass via $M_0 = 2\pi R_0 \mu$. The Schwarzschild radius of such configurations can then be expressed as
\begin{equation}\label{eq:Rbh}
    R_\mathrm{schw} \equiv \frac{2G M_0}{c^2} = 4\pi R_0 \frac{G\mu}{c^2}~.
\end{equation}
If during the loop's lifetime its mass is enclosed within its Schwarzschild radius, it will form a black hole \cite{Hawking:1987bn,Polnarev:1988dh,Caldwell:1993kv,Helfer:2018qgv,Bramberger:2015kua,James-Turner:2019ssu}. Using the hoop conjecture, this condition becomes rather generic when the loops are circular, as the initial radius needed to form a black hole is very small
\begin{equation}
    R_0 \gtrsim 10^{-26}  \,\mathrm{m}\,\left(\frac{G\mu/c^2}{10^{-7}}\right)^{-3/2}
\end{equation}
with masses
\begin{equation}\label{eq:mass}
    M_0 \approx 100\,M_\odot\left(\frac{R_0}{1\,\mathrm{AU}}\right)\left(\frac{G\mu/c^2}{10^{-7}}\right)\,.
\end{equation}
Here $1\,\mathrm{AU}\approx 1.49\times 10^8\,\mathrm{km}$ denotes astronomical units, so these solar-system-sized loops are very small when compared to the size of the cosmological network. 

The gravitational waveform of a circular cosmic string loop can be described in three stages \cite{Aurrekoetxea:2020tuw}: infall, black hole formation and ringdown:
\begin{equation}
h (t)=
\begin{cases}
h_\mathrm{[infall]} \qquad  t < t_\mathrm{bh}\,,\\
h_\mathrm{[bh]} \qquad  t_\mathrm{bh} < t < t_\mathrm{qnm}\,,\\
h_\mathrm{[qnm]} \qquad   t_\mathrm{qnm} < t\,.
\end{cases}    
\end{equation}
Analogous to BBHs~\cite{Khan:2015jqa,Bohe:2016gbl}, string loop waveforms can be constructed using semi-analytical and numerical techniques.\\

\noindent
\textit{Infall:} During the early stages of the loop collapse, when its radius is much larger than its Schwarzschild radius, local backreaction effects can be neglected and the dynamics can be treated within the weak-field limit,
\begin{equation}\label{eq:weak_grav}
h_{ij}^{[\mathrm{infall}]}(t) = \frac{4G}{r c^4}\int_{-\infty}^\infty d^3 x~T_{ij}\left(t_\mathrm{ret}+\mathbf{x}\cdot\mathbf{n},\mathbf{x} \right),
\end{equation}
where $t_\mathrm{ret}\equiv t-r/c$, and $r$ and $\mathbf{n}$ are the distance and direction of the observer \cite{maggiore2008gravitational}. In the above expression, $T_{ij}$ is the energy momentum tensor, which for a (infinitesimally thin) circular cosmic string loop oscillating on the $z=0$ plane is given by
\begin{equation}
T^{\alpha\beta}(t,\mathbf{x}) =  \mu\, \delta\left[\sqrt{x^2+y^2}-R(t)\right]\delta[z]\, U^{\alpha} U^{\beta},
\end{equation}
where $U^\alpha = \gamma \, (c,\, V(t)\sin(\phi),\, V(t)\cos(\phi),\, 0)$ is the four-velocity and $\gamma$ the Lorentz contraction factor. The radius and velocity during the infall are given by \cite{Nambu:1969se,Helfer:2018qgv}
\begin{align}\label{eq:R(t)} 
    R(t) &= R_0 \cos\left(\frac{c\,t}{R_0}\right), \quad V(t) = c \sin\left(\frac{c\,t}{R_0}\right)\,.
\end{align}
Even if the periodicity of these solutions seems to suggest the existence of ever-oscillating circular loops (via the trigonometric functions), this description breaks down when the loop radius is comparable to the string width. In this work we will focus on the last oscillation of a planar, circular loop, where the endpoint is the formation of a black hole. Geometrically, non-circular loops can still form black holes as long as the loops are planar. We expect non-planar and  non-self-intersecting loops will eventually circularize via radiation of scalars, vectors, and gravitational waves, but this timescale is not known.
The maximum radius of a circular loop $R_0$ sets the oscillation frequency and thus the frequency content of the infall signal, which is approximately
\begin{equation}\label{eq:f_infall}
    f_\mathrm{[infall]}\approx 10^{-3}\, \mathrm{Hz}\left(\frac{1\,\mathrm{AU}}{R_0}\right)\,.
\end{equation}
From Eqn. \eqref{eq:weak_grav}, it can be shown that, 
\begin{align}
    r h^+_{[\mathrm{infall}]}(t,\theta,\phi) &= R_0 \frac{G\mu}{c^2} \mathcal{I}(t,\theta,\phi) \\
    r h^\times_{[\mathrm{infall}]} (t,\theta,\phi) &= 0
\end{align}
where $h^\times_{[\mathrm{infall}]} = 0$ due to the axial symmetry of the collapse and the details of the numerical integral $\mathcal{I}(t,\theta,\phi)$ can be found in the appendices of Ref. \cite{Aurrekoetxea:2020tuw}.  A template bank can be constructed using the scaling relations
\begin{align}
    h^+_{[\mathrm{infall}]} &\propto \frac{R_0}{r}\frac{G\mu}{c^2} \propto \frac{1}{r}\frac{GM_0}{c^2}\,,\\
    \Delta t_{[\mathrm{infall}]}&\propto \frac{R_0}{c}\,.
\end{align}

\noindent
\textit{Black hole formation:} At early times, the weak-field GW description of the collapsing loop during its infall stage is valid. However, this linearized treatment breaks down when the loop's size is comparable to its Schwarzschild radius, and gravitational effects become significant. We define this time $t_\mathrm{bh}$, when $R(t_\mathrm{bh})\approx R_\mathrm{schw}$ and can be estimated via Eqns. \eqref{eq:Rbh} and \eqref{eq:R(t)}, to be
\begin{equation}
    t_\mathrm{bh} \approx \frac{R_0}{c}\arccos\left(4\pi\frac{G\mu}{c^2}\right)~.
\end{equation}
This quantity sets the timescale of the infall, after which the waveform $h^+_\mathrm{[infall]}$ is no longer valid and needs to be matched to the black hole formation stage $h^+_\mathrm{[bh]}$. This is inherently a strong-gravity phenomenon and thus relies on general relativistic simulations. These become prohibitive when there exist several physical scales that need to be resolved, as is the case of extreme mass ratio inspirals of black holes. In the cosmic string scenario, we would need to resolve the string width $\delta\approx 10^{-35}\mathrm{km}$ (for $G\mu/c^2=10^{-7}$), and the Schwarzschild radius of the black hole $R_\mathrm{schw}\approx 500\mathrm{km}$ (for GW190521).
In addition, the Lorentz contraction of the loop along the collapsing direction at the latest stages of the infall is $\gamma \propto 1/G\mu\approx 10^{11}$. This results in a separation of scales $\delta/R_\mathrm{schw}\ll 10^{-37}$, beyond the capabilities of adaptive mesh refinement techniques \cite{Radia:2021smk}. However, results from numerical relativity simulations of collapsing cosmic string loops with $G\mu\approx 10^{-3}$ and $\delta/R_\mathrm{schw}\approx 10^{-2}$ exhibit a featureless intermediate stage between the infall and the ringdown, and we will thus model $h^+_\mathrm{[bh]}$ as a simple time-interpolating function between both phases. We choose this interpolation such that the radiated energy in gravitational waves agrees with the balance between the initial loop and final mass of the remnant black hole. \\

\noindent
\textit{Ringdown:} Slightly after the black hole has formed $t_\mathrm{qnm} \gtrapprox t_\mathrm{bh} + 15 GM/c^3$, the remnant enters the ringdown stage, $h^+_\mathrm{[qnm]}$. This is modelled as a linear sum of damped sinusoids
\begin{equation}
    h^+_\mathrm{[qnm]}(t) = \frac{1}{r}\sum_{\ell m} A_{\ell m}\exp\left[{\iu \omega_{\ell m} \frac{c^3}{GM}t}\right]
\end{equation}
where $\omega_{\ell m}=\mathrm{Re}(\omega_{\ell m}) + \iu \mathrm{Im}(\omega_{\ell m})$ are a set of complex frequencies known as quasi-normal modes. The real and imaginary parts set the oscillation frequency and damping rates of the sinusoids and can be obtained from perturbative calculations \cite{Kokkotas:1999bd,Berti:2009kk}, whilst the amplitudes $A_{\ell m}$ depend on the strong-field regime details of the event itself. In our case, the remnant black hole is non-spinning and predominantly radiates in the $(\ell ,m)=(2,0)$ multipole.\\

The radiation of energy-flux generally results in a permanent displacement $\Delta h=h(\infty)-h(-\infty)$ so-called gravitational-wave memory \cite{Zeldovich:1974gvh,Braginsky1987,PhysRevLett.67.1486,PhysRevD.45.520,Jenkins:2021kcj}. The magnitude can be estimated using the Christodolou formula $\Delta h \sim 4G E_\mathrm{rad}/c^4r$,
where $E_\mathrm{rad}$ is the total energy radiated. From general relativistic simulations \cite{Aurrekoetxea:2020tuw} and geometrical arguments by Hawking \cite{Hawking:1990tx}, collapsing circular cosmic string loops are expected to radiate between $1\%-29\%$ of their initial mass in gravitational waves. The expected memory\footnote{The significant memory obtained in \cite{Aurrekoetxea:2020tuw} was due to the large anisotropic emission of matter, which is not the case for our waveforms here as $\delta\ll R_\mathrm{Schw}$.} for sources that feature in the LVK frequency band is then given by $\Delta h \lessapprox 10^{-27}\,\left(1\,\mathrm{Gpc}/r\right)$, so we will ignore it when constructing the waveforms.\\

\begin{figure*}[t!]
    \centering
    \includegraphics[width=\linewidth]{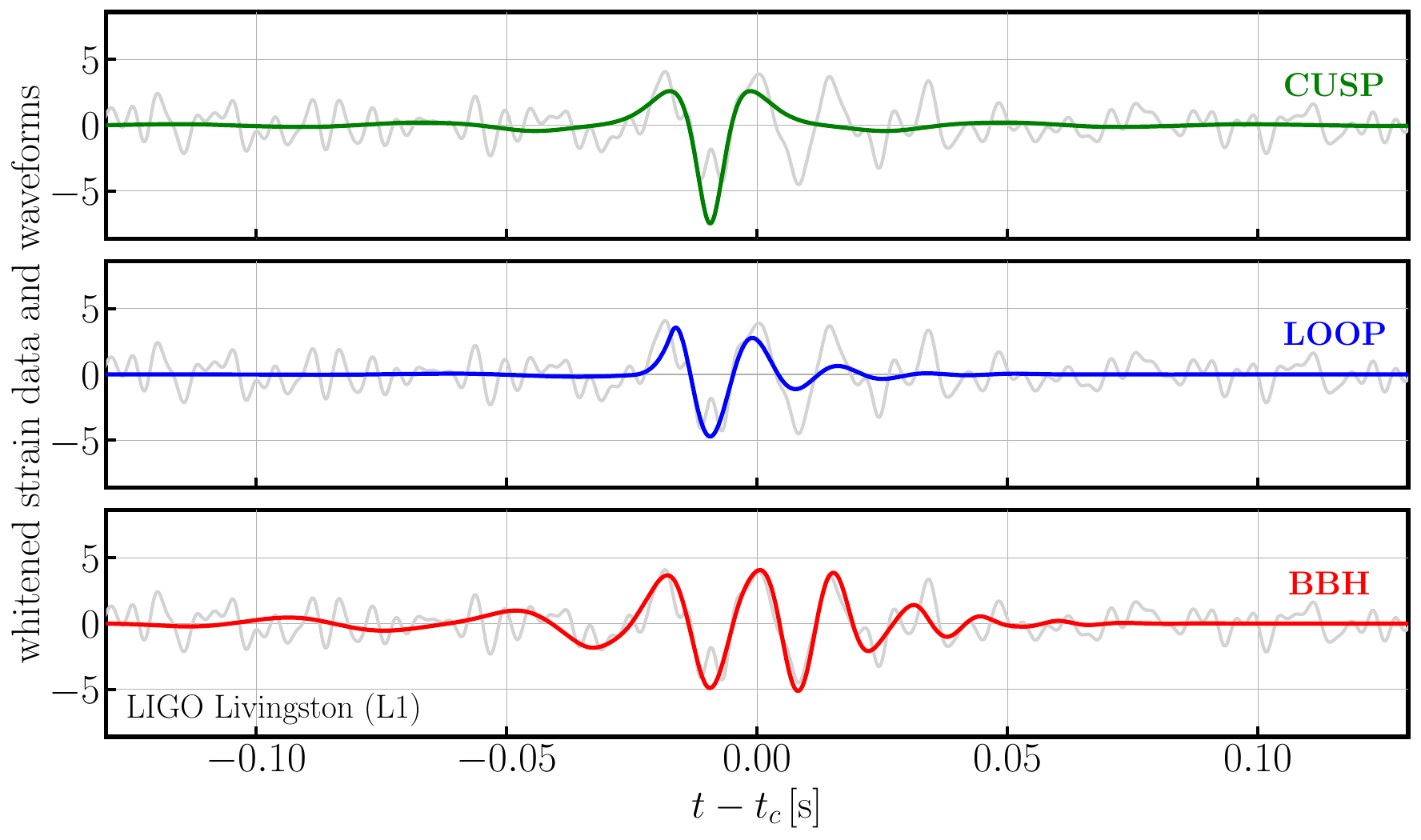}
    \caption{The time-domain gravitational wave strain data for the LIGO-Livingston detector~\cite{LIGOScientific:2014pky,LIGOScientific:2023vdi} around the time of GW190521~\cite{LIGOScientific:2020iuh}. The top, middle and bottom panels plot the best-fit detector-frame waveforms assuming a cosmic string cusp, loop collapse and  BBH merger respectively. The cosmic string cusp waveform is the best-fitting template from the LVK's cosmic string matched filter search pipeline~\cite{LIGOScientific:2017ikf,ligo_scientific_collaboration_and_virgo_2022_P2000158} and the BBH waveform is the reconstructed maximum likelihood waveform from the LVK analysis~\cite{LIGOScientific:2020iuh,ligo_scientific_collaboration_and_virgo_2022_6513631}. The strain data and waveforms are whitened by the noise power spectral density provided as part of the LVK data release~\cite{ligo_scientific_collaboration_and_virgo_2022_6513631}. We shift the data and waveforms by the coalescence time in the LIGO-Livingston detector $t_\mathrm{c}$ as reported in the LVK BBH analysis.}
    \label{fig:analysis}
\end{figure*}

\textbf{\textit{Data analysis.}}— Due to limited numerical relativity simulations for collapsing cosmic string loops and the difficulty of studying low string tensions, we do not construct a parameterized model, and hence are unable to perform a template based matched filter search or a Bayesian analysis to determine the parameters that maximises the signal-to-noise ratio (SNR). Instead, we calculated the matched-filter SNR for an example cosmic string loop with $G\mu/c^2=10^{-7}$ and $R_0=2.85\,\mathrm{AU}$ that radiates $3.4\%$ of its initial mass in gravitational waves. Although these may not correspond to the parameters that best fit GW190521, we demonstrate that our example waveform is still favoured when compared to previously studied cosmic string cusps.

The construction of waveforms is generally done in the source frame, whilst the analysis is in the detector frame. The transition between source-frame and detector-frame waveforms relies on additional parameters describing the sky position of the source, as well as its inclination $\iota$ and redshift $z$. When these sources are at cosmological distances, we infer a larger remnant mass in the detector frame $(1+z)M$ owing the redshifted waveforms. Similarly, the amplitude of the observed GWs is suppressed by a factor of $\sin^{2}\iota / d_{L}$ where $d_{L}$ is the luminosity distance to the source\footnote{Note that this is different from the expression used for BBHs, since the amplitude suppression depends on the multipole considered. BBHs predominantly emit in the ($\ell, m) = (2,2)$ while circular collapsing loops in the ($\ell, m) = (2, 0)$ multipole.}. As a result of the short timescales over which a significant fraction of the initial mass is radiated in gravitational waves, the amplitude of the waveform is large, see Fig. \ref{fig:waveform}. Suppressing the amplitude to typical GW strain levels restricts the allowed values of $\sin^{2}\iota / d_{L}$, and due to the degeneracy between $\iota$ and $d_{L}$, this can take any value between $0$ and $1/d_{L}$. We are therefore free to choose a distance and inclination angle. For this analysis, we use a fiducial distance $d_L=4\, \mathrm{Gpc}$, chosen to match the LVK BBH analysis~\cite{LIGOScientific:2020ufj}, and inclination angle $\iota = 34^\circ$. 

In Fig. \ref{fig:analysis} we plot the best-fit cusp \cite{LIGOScientific:2017ikf,ligo_scientific_collaboration_and_virgo_2022_P2000158} and loop waveforms with the strain data in the LIGO--Livingston GW detector. For comparison, we also add the the BBH waveform, which is the reconstructed maximum likelihood waveform from the LVK analysis~\cite{LIGOScientific:2020iuh,ligo_scientific_collaboration_and_virgo_2022_6513631}.
We whiten the data and waveforms using the publicly available power spectral densities (PSDs) provided as part of the LVK data release~\cite{ligo_scientific_collaboration_and_virgo_2022_6513631}. Owing to the short duration of GW190521 (approximately $0.1\,\mathrm{s}$), we additionally bandpass the strain data between $(10-256\, \mathrm{Hz})$ to suppress high frequencies. At early stages of the loop waveform, we see that the infall is restrained in comparison to Fig.~\ref{fig:waveform}. This is due to its low frequency nature, which is below the sensitivity of the LVK detectors~\cite{LIGOScientific:2014pky,acernese2014advanced,Somiya:2011np}. However, it retains the late-time ringdown, which is similar to the BBH waveform\footnote{The best-fit BBH waveform favours a spinning remnant, which has a lower damping rate of the ringdown. This allows it to capture more quasi-normal mode cycles than the non-spinning remnant black hole from the loop collapse.}. We see that the loop waveform is dominated by the black hole formation and ringdown stages, which solely depend on the mass of the loop (and thus black hole). This introduces a degeneracy between the radius and string tension via Eqn. \eqref{eq:Rbh}, which can only be broken in the presence of the infall signal as this timescale depends on $R_0$. This means that comparable waveforms with different string tensions and radii can be generated, as long as the mass of the loops remain unchanged.

\begin{figure}[t!]
    \centering
    \includegraphics[width=\linewidth]{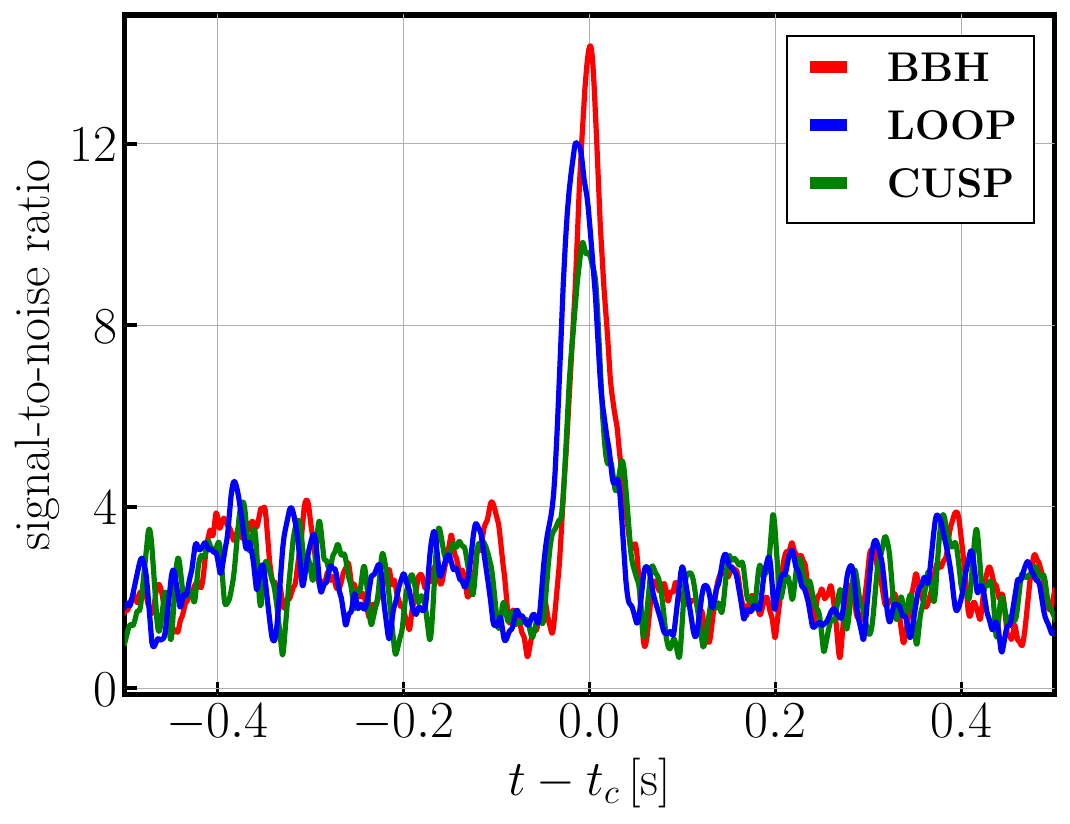}
    \caption{Network signal-to-noise ratios (assuming a network of the LIGO-Hanford, LIGO Livingston and Virgo detectors) for the best-fit BBH, loop and cusp waveforms, $\rho=\lbrace 14,\, 12,\, 10\rbrace$ respectively. As with Fig.~\ref{fig:analysis}, we shift the timeseries by the coalescence time in the LIGO-Livingston detector $t_\mathrm{c}$ as reported in the LVK BBH analysis.}
    \label{fig:SNR}
\end{figure}

To assess the probability that GW190521 was formed from a collapsing cosmic string, we calculate the matched-filter SNR for our example string loop waveform described above. This can be considered as analogous to a matched-filter search~\cite{Cannon:2011vi,
Privitera:2013xza,Messick:2016aqy,Hanna:2019ezx,
Sachdev:2019vvd,Usman:2015kfa,
Nitz:2017svb,Nitz:2018rgo,pycbc-software,2016CQGra..33q5012A,SPIIR2,
2012CQGra..29w5018L,2018CoPhC.231...62G} with only one template. For comparison, we repeat the calculation for the best-fit BBH and cusp models. We calculate the SNR $\rho$ by evaluating the noise-weighted inner product~\cite{Brown:2004vh}. We use the publicly available \texttt{PyCBC}~\cite{Usman:2015kfa,pycbc-software} software to perform the analysis. Often signal-consistency tests are additionally performed to distinguish between genuine astrophysical signals, and non-Gaussian noise artefacts~\cite{Davies:2020tsx}. We did not perform any signal-consistency tests in this work as we are ultimately interested in comparing the collapsing cosmic string, BBH and cusp models. In Fig. \ref{fig:SNR} we plot the SNR timeseries for our example collapsing cosmic string loop and best-fit BBH and cusp models. We see that the loop waveform greatly out performs the cusp model with an increase in SNR of $20\%$: $10$ to $12$. To highlight that our loop waveform provides a better fit to the data than previously considered cosmic string signals, we calculate the Bayes factor. We do not calculate the odds ratio, and therefore we do not account for the theoretical expectation of forming collapsing cosmic string loops in nature (a prior probability of the model), since their theoretical event rate is highly uncertain\footnote{See \cite{James-Turner:2019ssu} for details about constraining their collapse fraction via the abundance of primordial black holes. However, we note that cusps generically form, so we expect to have a much larger number of cusps in the network than circular loops (although collapsing loops are much louder events and can thus be observed from a larger volume).}. Assuming that the likelihood follows $\exp(-\rho^{2} / 2)$, the $\log$ Bayes' factor in favour of our loop waveform over cusps is $22$. This implies that the loop hypothesis is strongly preferred over the cusp model, and is therefore the most accurate representation of GW190521 assuming a cosmic string origin to date. We see that the BBH hypothesis obtains a larger SNR, meaning that it is still the preferred progenitor of GW190521. Without further details about the low-frequency nature of GW190521 and a more complete template bank of loop waveforms, we are unable of drawing further conclusions.\\

\textbf{\textit{Conclusion.}}— In this Letter we have revisited the origin of GW190521 using, for the first time, waveforms of circular cosmic string loops collapsing to black holes; a previously unconsidered source. We have shown that GW190521 can be described as collapsing string loop with a signal-to-noise ratio of $\rho \approx 12$. This greatly improves upon previous cosmic string cusp analyses, where $\rho\approx 10$ is obtained. By approximating the $\log$ Bayes' factor, we show that the collapsing cosmic string hypothesis is favoured over existing cusp results by $\log \mathcal{B}^{\mathrm{loop}}_{\mathrm{cusp}} = 22$.

Our cosmic string loop waveforms are constructed from circular, planar loops that have no angular momentum, and thus collapse to form Schwarzschild black holes. 
Indeed, this is a very special case -- in general loops will be non-planar \cite{Matsunami:2019fss}, or carry initial spin in its internal field configuration, or possess traveling kinks. A loop with initial angular momentum might collapse to form a black hole with significant spin. This would reduce the damping rate of the ringdown phase and improve the SNR, so it is entirely conceivable that such signals would be indistinguishable from binary black holes in an observation like GW190521. However, the details about the angular momentum loss rate during the loop collapse process and whether spinning black holes are to form from these systems is not known, so we leave this study for future work.

The timescales shown in Fig. \ref{fig:waveform} highlight the multi-frequency nature of these events. The data favours the absence of an oscillatory infall signal, suggesting that it must be below the low-frequency sensitivity of the LVK detectors, $f_\mathrm{[infall]}\lessapprox 10\,\mathrm{Hz}$. This implies that $R_0\gtrapprox 10^{-4}\,\mathrm{AU}$ via Eqn. \eqref{eq:f_infall}, and assuming the progenitor was a cosmic string loop of $M_0\approx 181 M_\odot$, we can put a weak constraint on the dimensionless string tension of this event to be $G\mu/c^2\lessapprox 10^{-2}$. This is not competitive with those obtained from CMB \cite{Lizarraga:2016onn} or SGWB \cite{LIGOScientific:2021nrg} searches, but note however that these are sensitive to details of the loop distribution model and how the string network evolves. 
Our work provides a novel technique to infer the astrophysical event rate of collapsing string loops by searching for their individual signals in LVK data. This could help guide the construction of string loop distribution models, analogous of how BBH detections help to constrain the BBH event rates and BH population models~\cite{KAGRA:2021duu}. Given that $\sim 80$ BBHs have been observed through GWs~\cite{LIGOScientific:2021djp}, and only one can be described by cosmic string loops with high significance, we derive a simple estimate for the event rate from collapsing loops to be $80$ times smaller than for BBHs. If circular loops were to exist and be detected with the methods described in this Letter, it could imply that either (i) there might be many more small loops than expected; and/or (ii) there exists a dynamical process that efficiently circularizes them, such as a significant radiation of gravitational waves.

Based on our results, we propose collapsing string loops as a potential hypothesis for future analyses. The degeneracy between high mass binary black hole and string collapse signals will make a definitive detection of this source difficult, although it might be possible to use higher modes to more conclusively distinguish them. The absence of additional power prior to black-hole formation may be a smoking gun in LVK observations, but it will be necessary to understand how easily string collapse can be distinguished from dynamical capture mergers that also lack significant pre-merger power. We suggest using ringdown-only waveform templates as a simple and viable approach to search for these events. Future GW detectors that target lower frequencies, such as LISA \cite{Auclair:2019wcv, Auclair:2023brk, LISAConsortiumWaveformWorkingGroup:2023arg}, will be able to probe the early-phase of GW events that feature at the edge of the LVK sensitivity, helping to distinguish between high mass BBHs or signals arising from new physics.

\textbf{\emph{Acknowledgements.}}— JCA would like to thank Thomas Helfer, Eugene Lim, Samaya Nissanke and Andrew Williamson for earlier collaboration in cosmic string simulations and discussions related to using higher modes to distinguish exotica. We also thank Jade Powell, Juan Calderon Bustillo, and Salvatore Vitale for comments on this manuscript. We are grateful for helpful discussions with Jose Juan Blanco-Pillado, Katy Clough, Pedro Ferreira, Ian Harry, Max Isi, Alex Jenkins, Laura Nuttall, and Ken Olum.
JCA acknowledges funding from the Beecroft Trust and The Queen's College via an extraordinary Junior Research Fellowship (eJRF). CH thanks the UKRI Future Leaders Fellowship for support through the grant MR/T01881X/1. MH thanks the Science and Technology Facilities Council (STFC) grant ST/V00154X/1.

Part of this work used the DiRAC@Durham facility managed by the Institute for Computational Cosmology on behalf of the STFC DiRAC HPC Facility (www.dirac.ac.uk) under DiRAC RAC15 Grant ACTP316. The equipment was funded by BEIS capital funding via STFC capital grants ST/P002293/1, ST/R002371/1 and ST/S002502/1, Durham University and STFC operations grant ST/R000832/1. DiRAC is part of the National e-Infrastructure.

This research has made use of data or software obtained from the Gravitational Wave Open Science Center (\href{https://www.gw-openscience.org}{https://www.gw-openscience.org}), a service of LIGO Laboratory, the LIGO Scientific Collaboration, the Virgo Collaboration, and KAGRA. LIGO Laboratory and Advanced LIGO are funded by the United States National Science Foundation (NSF) as well as the Science and Technology Facilities Council (STFC) of the United Kingdom, the Max-Planck-Society (MPS), and the State of Niedersachsen/Germany for support of the construction of Advanced LIGO and construction and operation of the GEO600 detector. Additional support for Advanced LIGO was provided by the Australian Research Council. Virgo is funded, through the European Gravitational Observatory (EGO), by the French Centre National de Recherche Scientifique (CNRS), the Italian Istituto Nazionale di Fisica Nucleare (INFN) and the Dutch Nikhef, with contributions by institutions from Belgium, Germany, Greece, Hungary, Ireland, Japan, Monaco, Poland, Portugal, Spain. KAGRA is supported by Ministry of Education, Culture, Sports, Science and Technology (MEXT), Japan Society for the Promotion of Science (JSPS) in Japan; National Research Foundation (NRF) and Ministry of Science and ICT (MSIT) in Korea; Academia Sinica (AS) and National Science and Technology Council (NSTC) in Taiwan. This material is based upon work supported by NSF's LIGO Laboratory which is a major facility fully funded by the National Science Foundation.

\emph{Software} -- We used Matplotlib~\cite{2007CSE.....9...90H} {\texttt{NumPy}}~\cite{harris2020array}, \texttt{pycbc}~\cite{pycbc-software} and {\texttt{SciPy}}~\cite{2020SciPy-NMeth} for analysis and plotting

\bibliography{main}

\end{document}